\documentclass[aps,prb,reprint,groupaddress]{revtex4-1} 
\usepackage{bm,braket,amsmath,mathtools,comment}
\usepackage{graphicx}
\usepackage{hyperref}

\begin{document}
\title{
Performance of the time-dependent variational principle for matrix product states in long-time evolution of a pure state
}
\author{Shimpei Goto\(^1\)}
\email[]{goto@phys.kindai.ac.jp}
\author{Ippei Danshita\(^{1,2}\)}
\email[]{danshita@phys.kindai.ac.jp}
\affiliation{\(^1\)Department of Physics, Kindai University, Higashi-Osaka, Osaka 577-8502, Japan} 
\affiliation{\(^2\)Yukawa Institute for Theoretical Physics, Kyoto University, Kyoto 606-8502, Japan} 
\date{\today}
\begin{abstract}
The projection of time-dependent variational principle (TDVP) for matrix product states enables us to perform long-time simulations of one-dimensional quantum systems with the conservation of the total energy and the norm of wave functions.
We compare long-time dynamics after a quantum quench simulated by TDVP with those by the exact diagonalization method in order to evaluate the performance of TDVP.\@
We show that in a nonintegrable model the projection of TDVP clearly improves the long-time behaviors of global observables included in the Hamiltonian, such as the kinetic and interaction energies. 
In contrast, this projection can lead to larger error for other observables than that caused by the truncation of states.
\end{abstract}
\maketitle
\section{Introduction\label{sec:introduction}}
Despite exponential increase of the dimension of the Hilbert space with system size, there are a few quantum many-body systems whose numerical simulations are feasible by classical computers.
A one-dimensional (1D) system is one of such numerically tractable quantum systems.
According to the area law, the entanglement entropy of energetically low-lying states of a gapped 1D system has an upper bound independent of the size of the system~\cite{hastings_area_2007,eisert_textitcolloquium_2010}.
These low entangled states can be efficiently described by matrix product states (MPSs) with not so large matrix dimensions, or bond dimensions~\cite{hastings_entropy_2007,schuch_entropy_2008}.
With numerical energy optimization methods based on MPS, namely density matrix renormalization group (DMRG), one can access static properties of these states efficiently
~\cite{white_density_1992,white_density-matrix_1993,mcculloch_density-matrix_2007,schollwock_density-matrix_2011}.

In contrast, it is in general intractable to simulate long time evolution of quantum many-body systems even at 1D.
For instance, in quench dynamics triggered by a sudden and substantial change of global parameters in the Hamiltonian, the creation of a number of excitations causes significant growth of the entanglement entropy.
In clean systems, it grows linearly with time~\cite{calabrese_evolution_2005,calabrese_entanglement_2009}, and thus bond dimensions grow exponentially with time~\cite{schollwock_density-matrix_2011}.
If one uses standard techniques of time evolution of MPS, such as time-evolving block decimation (TEBD)~\cite{vidal_efficient_2003,vidal_classical_2007,daley_time-dependent_2004,white_real-time_2004}, the Krylov subspace method~\cite{schmitteckert_nonequilibrium_2004,manmana_time_2005,garcia-ripoll_time_2006,rodriguez_coherent_2006,wall_out--equilibrium_2012}, and some methods based on the matrix product operator description~\cite{zaletel_time-evolving_2015,goto_cooling_2017}, such exponential growth forces one to truncate relevant states in MPS.\@
This truncation results in artificial changes of the total energy and the norm of wave function under unitary evolution by a time-independent Hamiltonian.
Though a recently proposed truncation scheme based on density matrix representation does not change the norm and the total energy of short-range Hamiltonians, there remains a problem that the truncated density matrix does not necessarily represent a pure state even if an initial density matrix represents a pure state~\cite{white_quantum_2018}.
For another approach based on the expansion of the time-evolution operator by Chebyshev polynomials, the finite-order truncation of the expansion limits reachable simulation time~\cite{halimeh_chebyshev_2015}.

Recent development of time-dependent variational principle (TDVP) for MPS~\cite{haegeman_time-dependent_2011,haegeman_post-matrix_2013,haegeman_unifying_2016} has opened up new possibilities for long-time simulations of quantum many-body systems.
Since the truncation of relevant states is not necessary in TDVP, it allows for the time evolution with the conservation of the total energy and the wave-function norm.
Thanks to this property, TDVP is expected to describe long-time behaviors of physical quantities better than the other methods mentioned above.
In recent studies, it has been examined whether TDVP can capture thermalizing dynamics starting with states at infinite temperature~\cite{leviatan_quantum_2017,kloss_time-dependent_2018} or a single product state in disordered systems\cite{doggen_many-body_2018}.
These studies have reported both positive and negative results: TDVP captures the long-time behaviors of nonintegrable spin chains while it fails in describing those of integrable chains or nonintegrable ladders.

In this paper, we investigate the capability of the TDVP in quench dynamics starting with a single pure state in clean systems, taking nonintegrable Bose-Hubbard and integrable Fermi-Hubbard Hamiltonians as specific examples.
We choose these models because they are regarded as fundamental models for interacting quantum many-body systems and their quench dynamics can be realized experimentally with ultracold gases in optical lattices~\cite{greiner_collapse_2002,schneider_fermionic_2012,trotzky_probing_2012,cheneau_light-cone-like_2012,meinert_observation_2014}. 
Unlike previous studies~\cite{leviatan_quantum_2017,kloss_time-dependent_2018,doggen_many-body_2018}, the samplings of initial states or disordered potentials are absent in the present study. 
This difference is important since samplings can help low entangled states to describe the expectation value of a highly entangled state. 
For instance, in the algorithm called minimally entangled typical thermal states~\cite{white_minimally_2009,stoudenmire_minimally_2010}, 
expectation values of observables at finite temperatures can be accurately computed by sampling many low entangled states.
If one wants to compute such expectation values with a single pure state, 
the pure state typically has to hold large entanglement obeying the volume law~\cite{eisert_textitcolloquium_2010}.
The present study focuses on the performance of the TDVP without the helps of such samplings.

By comparing results obtained by TDVP with those by the exact diagonalization (ED)~\cite{raventos_cold_2017}, we show that the conservation of the total energy and the wave-function norm by TDVP does not simply lead to more accurate time evolution of quantum states.
For quench dynamics of nonintegrable models, we find that the TDVP correctly captures the long-time behavior of global observables included in the Hamiltonian, such as total kinetic energy and total interaction energy, with small bond dimensions.
However, TDVP fails to describe other observables, which are not included in the Hamiltonian, and can give even worse description than the other time-evolution methods with the truncation of states.
These results mean that the time-evolved states obtained by TDVP are biased in favor of optimizing the total energy and lucks the capability to describe arbitrary observables.
On the other hand, for integrable models, we show that TDVP fails even in describing the global observables included in the Hamiltonian.
This observation is consistent with previous studies on other models at infinite temperature~\cite{kloss_time-dependent_2018} and can be attributed to the fact that TDVP does not respect nonlocal conserved quantities resulting from the integrability.

The rest of the paper is organized as follows:
In Sec.~\ref{sec:method}, we briefly explain TDVP algorithm and its variants.
Time-evolution scheme of MPS used in this paper is also introduced in this section.
In Sec.~\ref{sec:result}, results obtained by TDVP and ED are shown.
Comparing these results, we show that the energy conservation character of TDVP leads to some biases to time-evolved state.
Summaries are given in Sec.~\ref{sec:summary}.

\section{Time dependent variational principle: Two variants and their properties\label{sec:method}}
Any wave function \(\ket{\psi}\) on a \(L\)-site lattice system has a MPS representation
~\cite{schollwock_density-matrix_2011}
\begin{equation}
\label{eq:MPS}
\ket{\psi} = \sum_{\bm{\sigma}}\bm{A}^{\sigma_1}_1 \bm{A}^{\sigma_2}_2 \cdots \bm{A}^{\sigma_L}_L \ket{\bm{\sigma}},
\end{equation}
where \(\sigma_i\) is the state of the local Hilbert space at \(i\)-th site,
\(\ket{\bm{\sigma}} = \ket{\sigma_1, \sigma_2, \ldots, \sigma_L}\), 
and \(\sum_{\bm{\sigma}}\) means the summation over all possible configurations of \(\sigma_i\).
The matrix dimensions of matrices \(\bm{A}^{\sigma_i}_i\) are called bond dimensions.
When a system is parted into subsystems \(A\) and \(B\), 
the entanglement entropy of subsystem \(A\) is given by
\begin{equation}
  \label{eq:entent}
  S_A = -\mathrm{Tr} \left[\rho_A \ln \rho_A\right],
\end{equation}
where \(\rho_A\) is a reduced density matrix defined as
\begin{equation}
  \label{eq:red_den_mat}
  \rho_A = \mathrm{Tr}_B \ket{\psi} \bra{\psi}.
\end{equation}
Here, \(\mathrm{Tr}_B\) means a partial trace over subsystem \(B\)
~\cite{schollwock_density-matrix_2011}.
If we divide the system at the link between sites \(i\) and \(i+1\),
 the bond dimension of \(\bm{A}^{\sigma_i}_i\) should be larger than \(\exp S_A\) in order to represent the state faithfully~\cite{schollwock_density-matrix_2011}.

In the TDVP scheme, the Schr\"odinger equation for the system described by the Hamiltonian \(\hat{H}\),
\begin{equation}
  \label{eq:schrodinger}
  \mathrm{i} \hbar \frac{\partial}{\partial t} \ket{\psi(t)} = \hat{H} \ket{\psi(t)}
\end{equation}
is projected to the manifold of a MPS representing \(\ket{\psi(t)}\)~\cite{haegeman_time-dependent_2011,haegeman_post-matrix_2013,haegeman_unifying_2016}.
In other words, instead of Eq.~\eqref{eq:schrodinger}, we solve a projected Schr\"odinger equation
\begin{equation}
  \label{eq:tdvp}
  \mathrm{i} \hbar \frac{\partial}{\partial t} \ket{\psi(t)} = \mathcal{P}_M \hat{H} \ket{\psi(t)}.
\end{equation}
Here, \(\mathcal{P}_M\) is a projector to the manifold of a MPS.\@

Recently, Haegeman \textit{et al}.~have introduced a very useful projection scheme and shown that TDVP can be implemented by replacing the diagonalization part of the DMRG procedure with the matrix exponential action of projected effective Hamiltonian (and some additional procedures for gauging a MPS)\cite{haegeman_unifying_2016}
likewise a time-step targeting method~\cite{feiguin_time-step_2005}.
The DMRG algorithms consist of successive optimizations of local MPS matrices \(\bm{A}^{\sigma_i}_i\) to minimize the energy of a system, and thus one can devise some variants of the DMRG by changing the number of sites to be optimized at one update.
The traditional DMRG algorithm~\cite{white_density_1992,white_density-matrix_1993} adopts the two-site update scheme.
Since updating only one site is more numerically efficient than updating two sites,
some DMRG algorithms based on one-site update have been invented~\cite{white_density_2005,hubig_strictly_2015}.
In these one-site update schemes, however, there is a drawback from one-site nature: the basis of matrices \(\bm{A}^{\sigma_i}_i\) cannot be changed and the energy is easily stacked at a local minimum.
In order to overcome this drawback,
these one-site DMRG algorithms include procedures to expand the basis of \(\bm{A}^{\sigma_i}_i\) with utilizing noisy effects.

On the basis of the close similarity in implementation between TDVP and DMRG, two variants of the TDVP scheme have been developed, namely one-site and two-site integration schemes~\cite{haegeman_unifying_2016}.
The one-site integration scheme does not change the basis of matrices \(\bm{A}^{\sigma_i}_i\) and thus the bond dimensions do not increase during an integration.
Thanks to this fixed bond dimensions, the truncation of states is not required so that the time evolution does not violate the conservation of the total energy and the wave-function norm.
A compensation for the fixed bond dimensions is an error coming from the projection \(\mathcal{P}_M\) to the manifold of MPS.\@
On the other hand, in the two-site integration scheme,
the basis of \(\bm{A}^{\sigma_i}_i\) changes during an integration and the bond dimensions generally increase.
This means that one has to truncate states during the time evolution, which violates the conservation, in order to avoid the exponential growth of the bond dimensions with time.
A main advantage of increasing bond dimensions is the absence of the projection error for 1D nearest-neighbor Hamiltonians~\cite{haegeman_unifying_2016}.

In this work, in order to take the advantages of the two schemes and diminish their shortcomings, we use TDVP with a simple hybrid scheme.
In quench dynamics, an initial state is a low-energy eigenstate of a certain Hamiltonian.
In a 1D system, this initial state can be represented by MPS with small bond dimensions thanks to the area law.
At an early stage of the time evolution, while the required size of the bond dimensions grows gradually with time, it is still modest so that simulations with classical computer can track the exact dynamics. 
For such early-stage dynamics, we use the two-site integration scheme. 
When the largest bond dimension reaches a certain threshold \(M_{\rm th}\), we switch from the two-site scheme to the one-site scheme.
After this switching, the bond dimensions do not increase any more and the total energy of the system is conserved.
Besides, since a MPS used in the projection \(\mathcal{P}_M\) has relatively large bond dimensions determined by \(M_{\rm th}\),
the projection error due to the one-site scheme is expected to be smaller than the case in which the one-site scheme is used solely.

The implementation of the TDVP is based on Ref. \onlinecite{haegeman_unifying_2016} and the Krylov subspace method is used for calculating the matrix exponential actions of local effective Hamiltonians~\cite{hochbruck_krylov_1997,moler_nineteen_2003,wang_error_2017}.
We also use the Krylov method in the ED based method to calculate the exponential actions~\cite{park_unitary_1986,manmana_time_2005}.
The entanglement entropy is calculated from the singular value decomposition of a wave function obtained by MPS based~\cite{schollwock_density-matrix_2011} or ED based~\cite{alba_boundary-locality_2012} simulations.
In the two-site integration scheme, we set the bond dimensions in such a way that the truncation error is smaller than \(10^{-10}\) or set them to be \(M_{\rm th}\) when the truncation error exceeds \(10^{-10}\).

Using the procedure described above, we evaluate the performance of the TDVP schemes via simulations of long-time quench dynamics of the 1D extended Bose-Hubbard model,
\begin{align}
  \label{eq:BoseHubbard}
  \begin{aligned}
  \hat{H}^B &= \hat{H}^B_0 + \hat{H}^B_\mathrm{int}, \\
  \hat{H}^B_0 & = -J \sum_{i} (\hat{b}^\dagger_i \hat{b}_{i+1} + \mathrm{H.c.}), \\
  \hat{H}^B_\mathrm{int} & = \frac{U}{2} \sum_{i} \hat{n}^B_i (\hat{n}^B_i - 1) + V \sum_{i} \hat{n}^B_i \hat{n}^B_{i+1},
  \end{aligned}
\end{align}
which is nonintegrable,
and the Fermi-Hubbard model with a staggered magnetic field,
\begin{align}
  \label{eq:FermiHubbard}
  \begin{aligned}
	  \hat{H}^F &= \hat{H}^F_0 + \hat{H}^F_\mathrm{int} + \hat{H}^F_\mathrm{stagg}, \\
  \hat{H}^F_0 &= -J \sum_{i \sigma} (\hat{c}^\dagger_{i \sigma} \hat{c}_{i+1 \sigma} + \mathrm{H.c.}), \\
  \hat{H}^F_\mathrm{int} &= U \sum_{i} \hat{n}^F_{i\uparrow} \hat{n}^F_{i\downarrow} \\
  \hat{H}^F_\mathrm{stagg} &= h\sum_i {(-1)}^i (\hat{n}^F_{i\uparrow} - \hat{n}^F_{i\downarrow}).
  \end{aligned}
\end{align}
which is integrable in the absence of the staggered field.
Here, \(J\) is the hopping amplitude, \(U\) is the on-site Hubbard interaction, \(V\) is the nearest-neighbor interaction,
\(\hat{b}_i\) \((\hat{b}^\dagger_i)\) annihilates (creates) a boson at site \(i\),
\(\hat{n}^B_i = \hat{b}^\dagger_i \hat{b}_i\),
\(\hat{c}_{i\sigma}\) \((\hat{c}^\dagger_{i\sigma})\) annihilates (creates) a fermion with spin \(\sigma \) at site \(i\),
and \(\hat{n}^F_{i\sigma} = \hat{c}^\dagger_{i \sigma} \hat{c}_{i \sigma}\).
The staggered field \(h\) is added for preparing the N\'eel state as a simple initial state and the time evolution shown in the next section is performed at the integrable point, \(h = 0\). 
The time-step size during time evolution is dynamically changed up to \(0.05 \hbar J^{-1}\) in order to efficiently obtain data on a logarithmic time scale.
For the Bose-Hubbard model, we set the maximum occupation number of bosons per site to be ten throughout the paper. 
Notice that the ground-state phase diagrams of the two models have been previously revealed in a broad parameter region by means of analytical and accurate numerical methods~\cite{kuhner_one-dimensional_2000,pai_superfluid_2005,batrouni_supersolid_2006,berg_rise_2008,lieb_absence_1968}. 

\section{Comparison with exact numerical data\label{sec:result}}
Firstly, we investigate time evolution for the Bose-Hubbard model \(\hat{H}^B\) with \(U/J = 3.01\) and \(V/J=0\).
The system size \(L\) and the total particle number \(N\) are set to \(L=N=14\), at which quench dynamics can be computed with the ED based method.
As an initial state of time evolution, we choose either of the following two states.
The first one is a Mott insulating state at unit filling in the atomic limit represented by a classical product state \(\ket{\psi} = \prod_i b^\dagger_i\ket{0}\), where \(\ket{0}\) is the vacuum state.
The choice of this initial state corresponds to the sudden change of the parameter \(U/J\) from \(\infty \) to \(3.01\).
The second one is the ground state of the noninteracting Hamiltonian \(\hat{H}^B_0\) at unit filling.
The parameter \(U/J=3.01\), at which the ground state is in a superfluid phase near the quantum critical point, is chosen so that the total energy of the state quenched from \(U/J=\infty \) is close to that of the state from \(U/J = 0\).

\begin{figure}
  \includegraphics[scale=0.4]{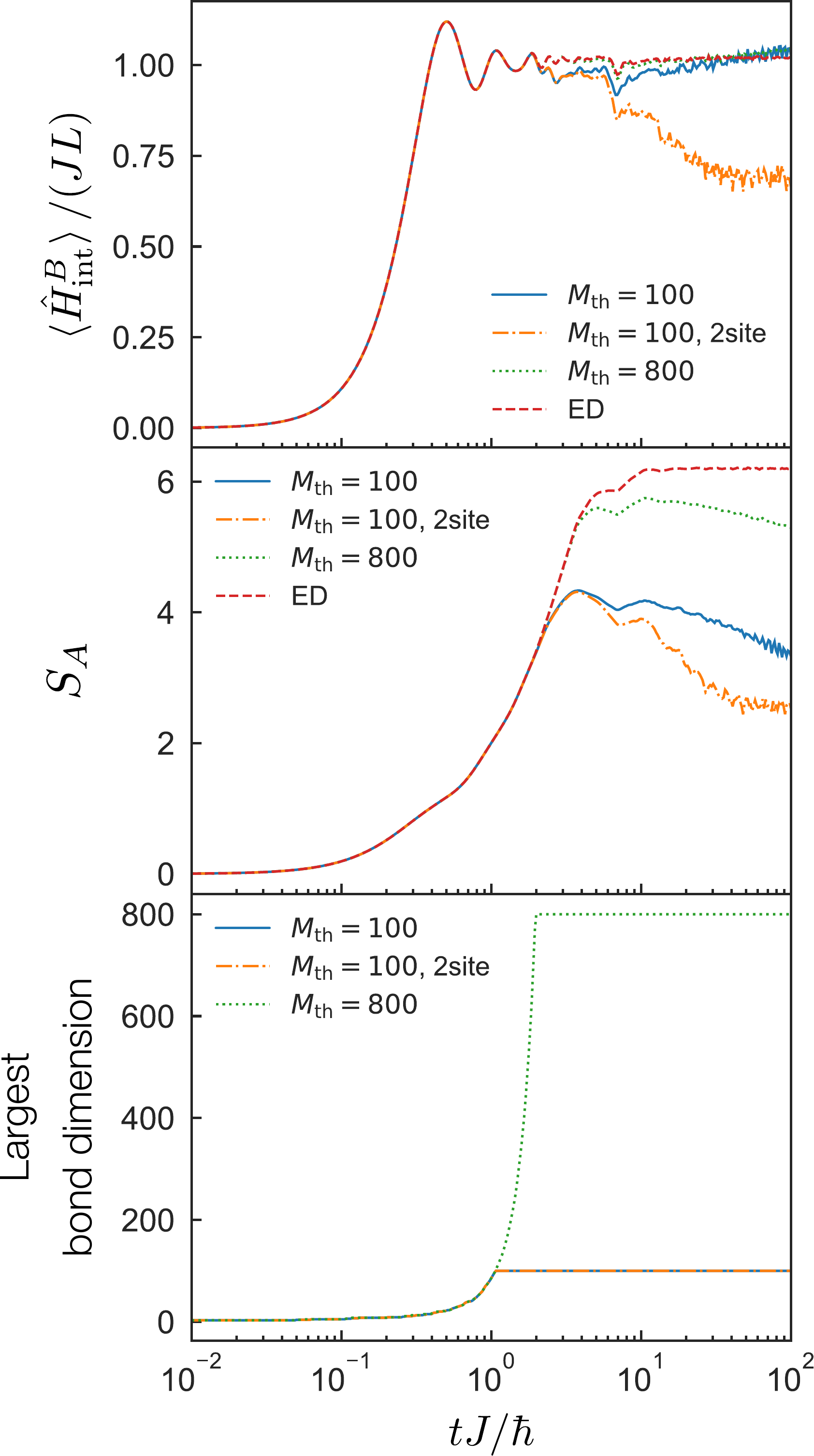}
  \caption{(Color online) Time evolution of the three quantities for the Bose-Hubbard model after the global quench from \(U/J=\infty \) to 3.01 with different integration schemes, where \(L=N=14\) and \(V/J = 0\). Top panel: Interaction energy \(\braket{\hat{H}^{B}_\mathrm{int}}\).
  Middle panel: Entanglement entropy \(S_A\), where subsystem \(A\) is the left half of the system. Bottom panel: Largest bond dimension.
The blue solid and green dotted lines represent the results by the hybrid TDVP scheme with \(M_\mathrm{th}=100\) and \(800\).
The orange dash-dotted line represents the result by the TDVP with two-site integration scheme with \(M_\mathrm{th} = 100\). The red dashed line represents the ED scheme.\label{fig:BH_ene_inf}}
\end{figure}

\begin{figure}
  \includegraphics[scale=0.4]{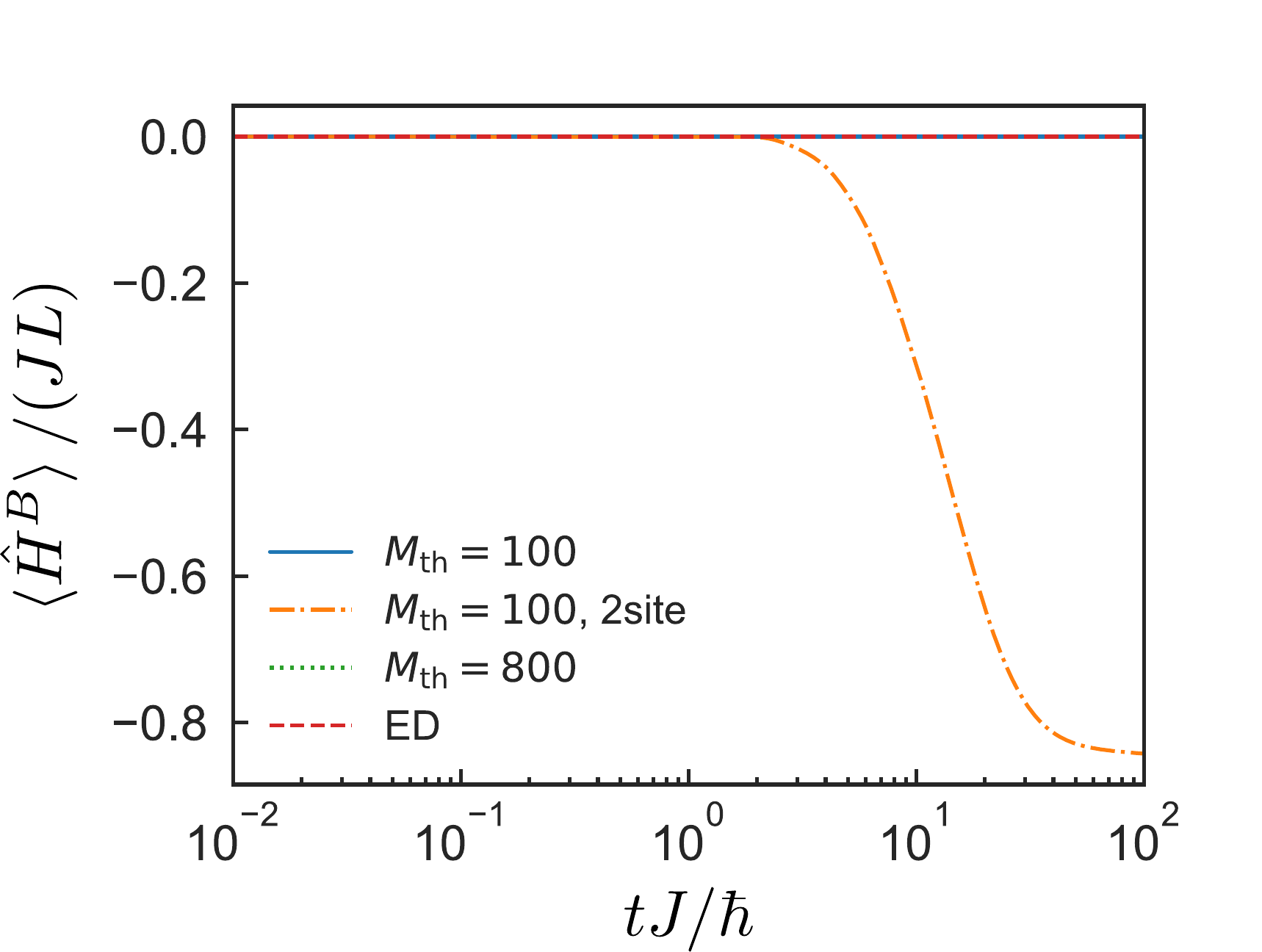}
  \caption{(Color online) Time evolution of the total energy for the Bose-Hubbard model \(\braket{\hat{H}^B}\) after the global quench from \(U/J = \infty \) to 3.01 with different integration schemes, where \(L=N=14\) and \(V/J = 0\).\label{fig:BH_tot_inf}}
\end{figure}

The top panel of Fig.~\ref{fig:BH_ene_inf} shows the time evolution of the interaction energy \(\braket{\hat{H}^B_\mathrm{int}}\) after the global quench of the onsite interaction from \(U/J = \infty \) to 3.01 with different integration schemes. Notice that the interaction energy of Hubbard-type models can be measured in ultracold atoms in optical lattices by means of the high-resolution spectroscopy of the local-atom-number distribution~\cite{takasu_experimental_2018}. 
For the same value of the threshold bond dimension, \(M_\mathrm{th} = 100\),
the hybrid scheme gives more accurate results than those given by the two-site scheme.
From the entanglement entropy shown in the middle panel of Fig.~\ref{fig:BH_ene_inf},
we see that the state evolved by the hybrid scheme is slightly more entangled than that by the two-site scheme.
In the viewpoint of the entanglement, the hybrid scheme also gives more accurate time-evolved states.

\begin{figure}
  \includegraphics[scale=0.4]{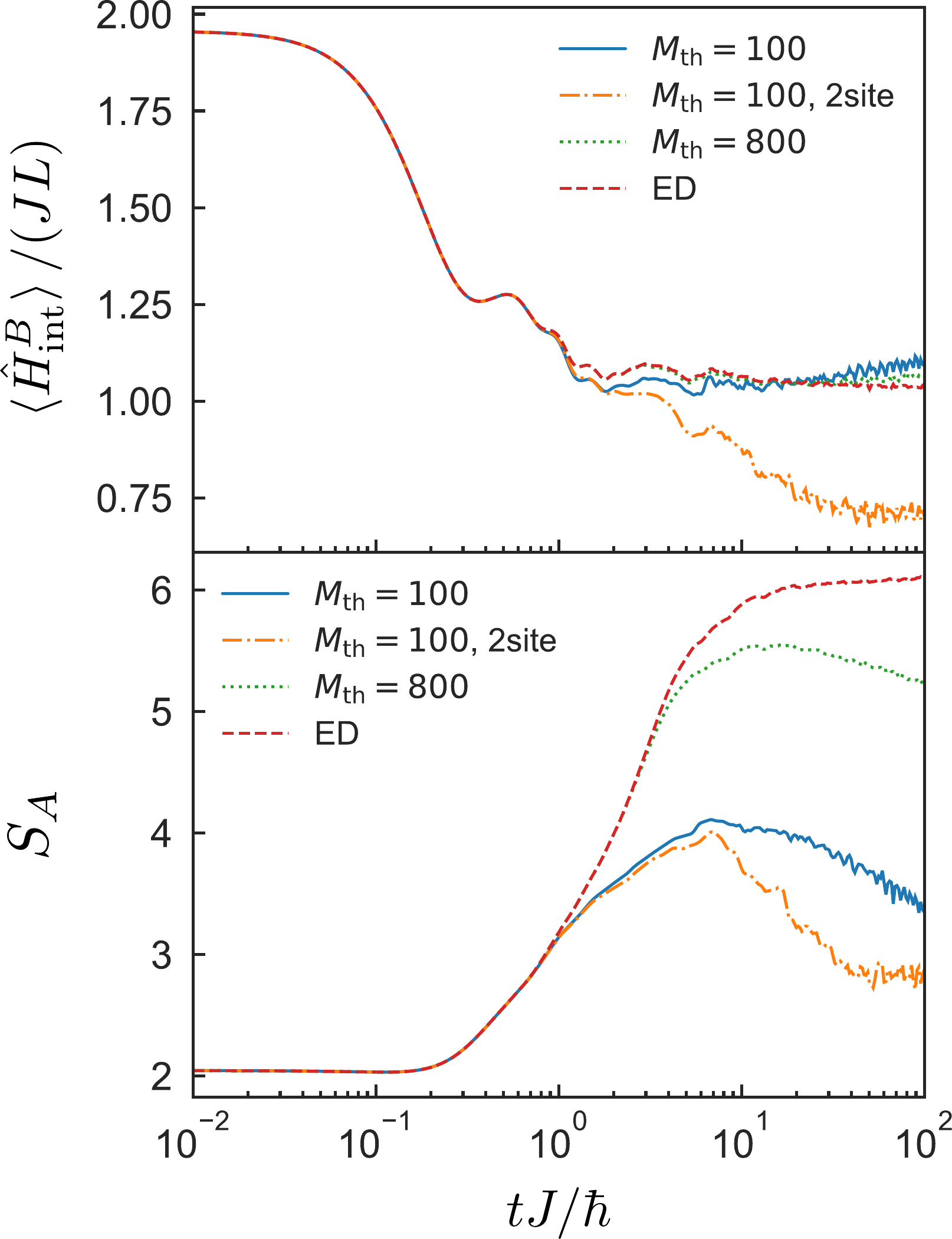}
  \caption{(Color online) Time evolution of the two quantities for the Bose-Hubbard model after the global quench from \(U/J = 0\) to 3.01, where \(L=N=14\) and \(V/J = 0\). Upper panel: Interaction energy \(\braket{\hat{H}^{B}_\mathrm{int}}\).
Lower panel: Entanglement entropy \(S_A\), where subsystem \(A\) is the left half of the system.}\label{fig:BH_ene_free}
\end{figure}

By comparing the results obtained by the hybrid scheme with \(M_\mathrm{th} = 100\) and \(800\), we clearly see that increasing \(M_\mathrm{th}\) reduces the projection error of the one-site integration scheme (See the bottom panel of Fig.~\ref{fig:BH_ene_inf}). 
The energy conservation property of the one-site integration scheme is confirmed from Fig.~\ref{fig:BH_tot_inf}, which depicts the time evolution of the total energy.
In the global quench of the Hubbard interaction from \(U/J = 0\) to 3.01,
we again observe the superiority of the hybrid scheme in describing the interaction energy and the entanglement entropy as shown in Fig.~\ref{fig:BH_ene_free}.

\begin{figure}
  \includegraphics[scale=0.4]{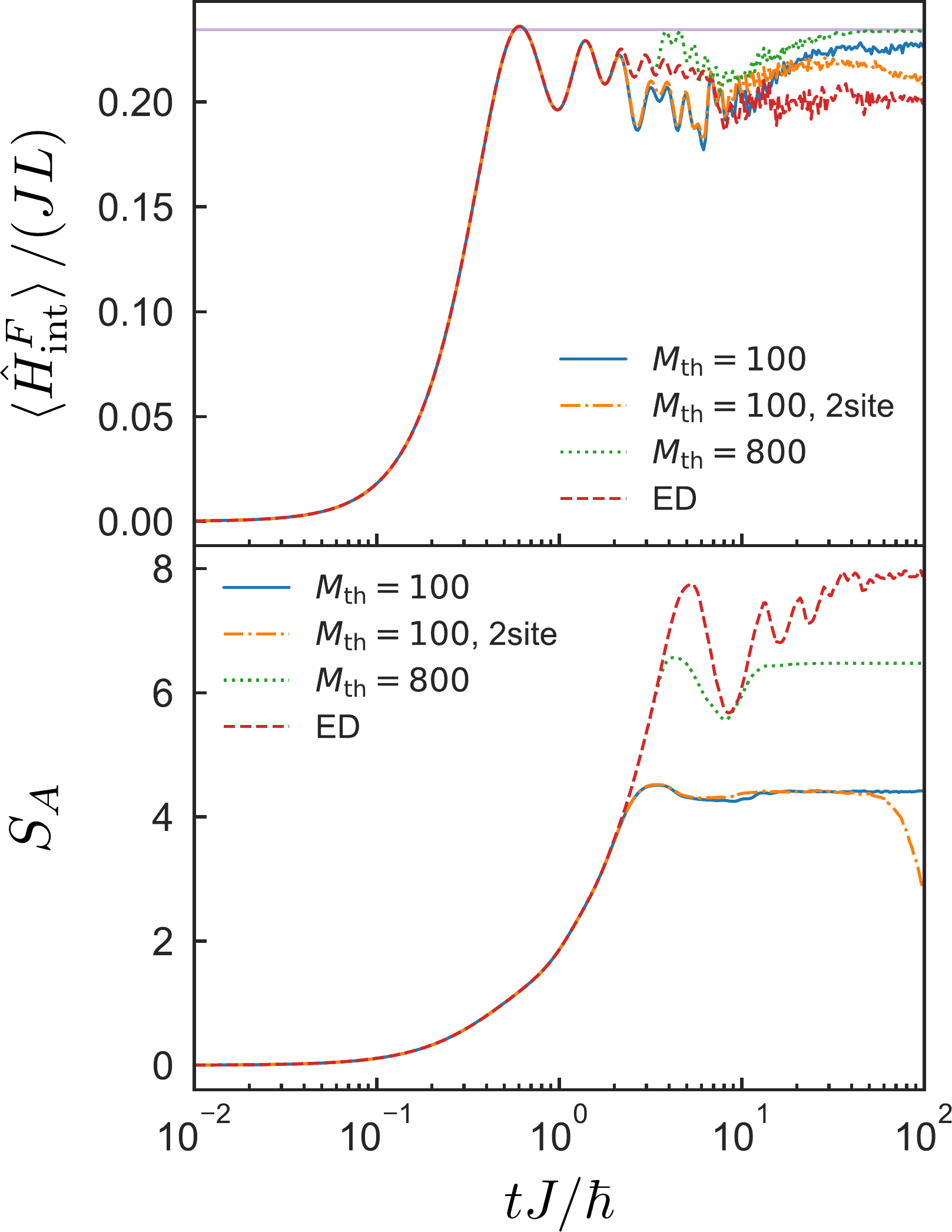}
  \caption{(Color online) Time evolution of two quantities for the Fermi-Hubbard model with \(U/J = 1.0 \) when we take the N\'eel state as an initial state, where \(L=14\) and \(N_{\uparrow} = N_{\downarrow} = 7\). Upper panel: Interaction energy \(\braket{\hat{H}^F_{\rm int}}\). The horizontal purple line represents the statistical expectation value given by a grand canonical ensemble. Lower panel: Entanglement entropy \(S_A\), where subsystem \(A\) is the left half of the system.}\label{fig:FH_ene_int}
\end{figure}

Next, we turn our attention to the integrable case of the Fermi-Hubbard model \( \hat{H}^F\).
We take the N\'eel state \(\prod^{L/2}_{i=1} c^\dagger_{2i-1 \uparrow} c^\dagger_{2i \downarrow} \ket{0}\), which is the ground state of the staggered Fermi-Hubbard model at \(h = \infty \), as an initial state of quench dynamics.
The upper panel of Fig.~\ref{fig:FH_ene_int} shows the time evolution of the interaction energy for the Fermi-Hubbard model with \(U/J = 1.0\), \(L=14\), and \(N_{\uparrow} = N_{\downarrow} = 7\), where \(N_{\sigma}\) denotes the number of particles of spin \(\sigma \).
In contrast to the nonintegrable case, the interaction energy calculated by the hybrid scheme considerably deviates from the exact values in a long-time region, say, \(10 < tJ/\hbar < 100\).
For the entanglement entropy shown in the lower panel of Fig.~\ref{fig:FH_ene_int}, the superiority of the hybrid scheme is almost absent till around \(tJ/\hbar = 50\).

This deviation can be attributed to the fact that the time evolution by TDVP does not respect the nonlocal conserved quantities regarding the integrability.
Due to the presence of such nonlocal conserved quantities, local observables in integrable models at the long-time relaxation obey the generalized Gibbs ensemble~\cite{rigol_relaxation_2007,cassidy_generalized_2011,vidmar_generalized_2016},
\begin{equation}
  \hat{\rho}_\mathrm{GGE} = \frac{\exp(-\sum_k \lambda_k \hat{I}_k)}{\mathrm{Tr} \exp(-\sum_k \lambda_k \hat{I}_k)}.
\end{equation}
The GGE is characterized by the expectation values of the integrals of motion \(\hat{I}_k\) which are generally given by nonlocal operators~\cite{rigol_relaxation_2007,cassidy_generalized_2011,vidmar_generalized_2016}.
Coefficients \(\lambda_k\) are determined so that statistical values \(\mathrm{Tr} [\hat{I}_k \hat{\rho}_\mathrm{GGE}]\) give initial expectation values \(\braket{I_k}\).
Although these expectation values have to be conserved, the local update character of the TDVP does not respect the conservation of the integrals, except the total energy whose corresponding operator is \(\hat{H}^F\) and quantities protected by symmetries installed in the structure of a MPS~\cite{mcculloch_density-matrix_2007,schollwock_density-matrix_2011}, i.e.,
the total number of particles \(\hat{N} = \sum_{i\sigma} \hat{n}^F_{i\sigma}\) in this study.
With only the two integrals and setting \(\hat{I}_1 = \hat{H}^F, \lambda_1 = \beta, \hat{I}_2 = \hat{N}\) and \(\lambda_2 = - \beta \mu \)
, the GGE reduces to an ordinary grand canonical ensemble,
\footnote{Strictly speaking, there is one more integral \(\hat{I}_3 = \hat{M} = \sum_i (\hat{n}^F_{i\uparrow} -\hat{n}^F_{i\downarrow})\). However, a corresponding coefficient \(\lambda_3\) is 0 because of the spin-rotational symmetry and thus this integral is not shown in Eq.~\eqref{eq:GC}.}
\begin{equation}
\label{eq:GC}
\hat{\rho}_\mathrm{GC} = \frac{\exp[-\beta(\hat{H}^F - \mu \hat{N})]}{\mathrm{Tr}\exp[-\beta(\hat{H}^F - \mu \hat{N})]}.
\end{equation}
As shown in Fig.~\ref{fig:FH_ene_int}, the interaction energy computed by the one-site TDVP scheme indeed tends to relax towards the equilibrium value of the grand canonical ensemble, which is represented as the horizontal line.
Notice that for calculating the statistical expectation value \(\mathrm{Tr}[\hat{H}^F_\mathrm{int} \hat{\rho}_\mathrm{GC}]\),
we use the purification algorithm~\cite{feiguin_finite-temperature_2005,goto_cooling_2017} with setting an inverse temperature \(\beta \) to \(0.2613 J^{-1}\) and a chemical potential \(\mu \) to \(U/2\).
With these parameters, the internal energy \(\mathrm{Tr}[\hat{H}^F \hat{\rho}_\mathrm{GC}]\) is \(4.9 \times 10^{-4} J\) (\(\braket{\hat{H}^F} = 0\) for the N\'eel state) and the particle-hole symmetry assures \(\mathrm{Tr} [\hat{N} \hat{\rho}_\mathrm{GC}] = L\).

\begin{figure}
  \includegraphics[scale=0.4]{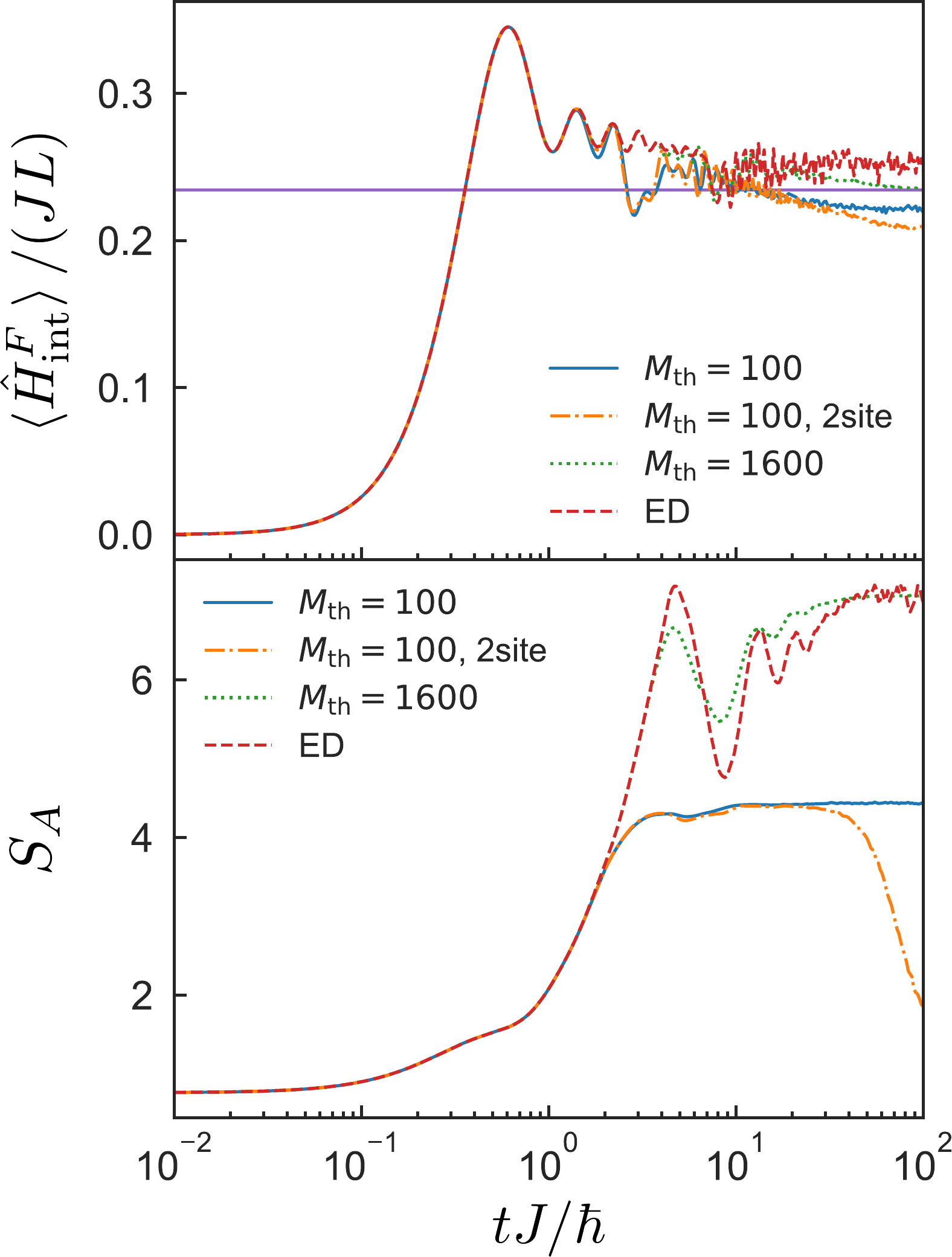}
  \caption{(Color online) Time evolution of two quantities for the Fermi-Hubbard model after the global quench from \(U/J = \infty \) to 1.0, where \(L=14\) and \(N_{\uparrow} = N_{\downarrow} = 7\). Upper panel: Interaction energy \(\braket{\hat{H}^F_{\rm int}}\). The horizontal purple line represents the statistical expectation value given by a grand canonical ensemble. Lower panel: Entanglement entropy \(S_A\), where subsystem \(A\) is the left half of the system.}\label{fig:FH_inf}
\end{figure}

\begin{figure}
  \includegraphics[scale=0.4]{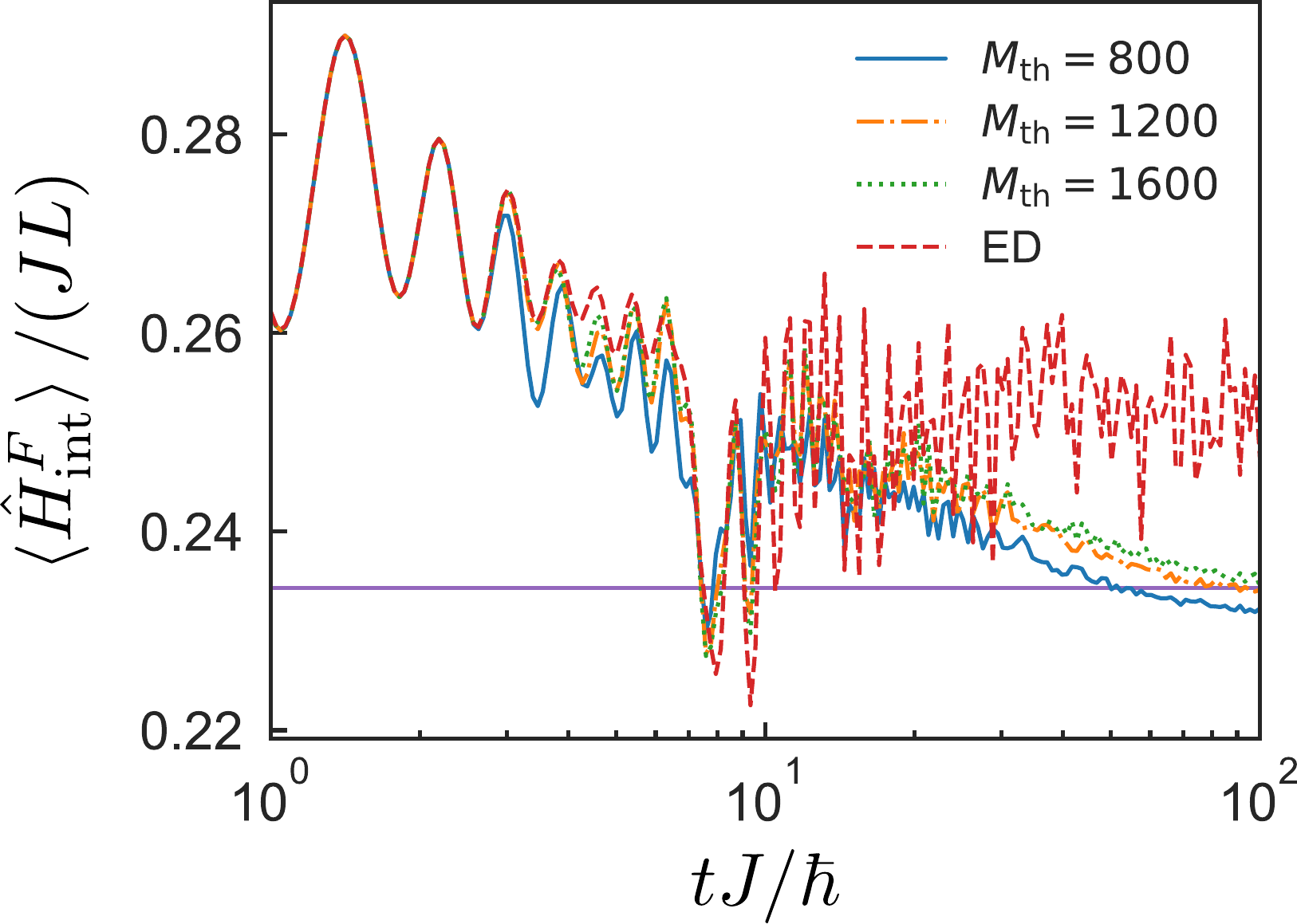}
  \caption{(Color online) Time evolution of interaction energy \(\braket{\hat{H}^F_{\rm int}}\) for the Fermi-Hubbard model after the global quench from \(U/J = \infty \) to 1.0 with larger threshold bond dimensions \(M_\mathrm{th} = 800, 1200\), and 1600, where \(L=14\) and \(N_{\uparrow} = N_{\downarrow} = 7\). The horizontal purple line represents the statistical expectation value given by a grand canonical ensemble.}\label{fig:FH_larger}
\end{figure}

In order to check how the choice of initial states affects the above explanation, we simulate another quench dynamics in the Fermi-Hubbard model.
As another initial state, we take the ground state of the Fermi-Hubbard model with \(U/J=\infty \), \(L= 14\) and \( N_\uparrow = N_\downarrow = 7\), i.e., the entangled ground state of the Heisenberg model,
\begin{align}
  \hat{H}^F_\infty &= \frac{4J^2}{U}\sum_{i}\left[\frac{1}{2}\left( \hat{c}^\dagger_{i\uparrow}\hat{c}_{i\downarrow}\hat{c}^\dagger_{i+1 \downarrow}\hat{c}_{i+1 \uparrow} + \mathrm{H.c.}\right) \nonumber \right. \\   &+ \left. \frac{1}{4}\left(\hat{n}_{i \uparrow} - \hat{n}_{i \downarrow}\right) \left(\hat{n}_{i+1 \uparrow} - \hat{n}_{i+1 \downarrow}\right)\right].
\end{align}
The total energy given by this ground state at \(U/J = \infty \) is also \(\braket{\hat{H}^F} = 0\) which is identical to the previous N\'eel state case.
The upper panel of Fig.~\ref{fig:FH_inf} represents the time-evolution of the interaction energy for the Fermi-Hubbard model after the global quench from \( U/J = \infty \) to 1.0.
Since the Fermi-Hubbard model is integrable, the interaction energy given by the ED relaxes to a different value from the relaxed value of the previous N\'eel-state case even though the total energy is the same.
In contrast, the relaxed value given by the hybrid scheme is very close to the value estimated from the ground canonical ensemble likewise the N\'eel-state case.
Furthermore, as shown in Fig.~\ref{fig:FH_larger}, this value seems to converge for the threshold bond dimension \(M_\mathrm{th}\).
It should be noted that the entanglement entropy of the largest \(M_\mathrm{th}\) case, say \(M_\mathrm{th} = 1600\), is comparable to that given by the ED as shown in the lower panel of Fig.~\ref{fig:FH_inf}.
These facts strongly support our explanation about the failure in the integrable model: The breaking of the nonlocal integrables of motion by the projection of the TDVP leads to a wrong thermalized value.

One may naively expect that the above discussion also gives the explanation for the success of the hybrid scheme in the nonintegrable model shown in Figs.~\ref{fig:BH_ene_inf} and~\ref{fig:BH_ene_free}.
In other words, the one-site TDVP scheme can capture the relaxation towards the equilibrium value of the grand canonical ensemble, which local observables of nonintegrable models obey in general, because it  respects the conservation of the total energy and the total number.

\begin{figure}
  \includegraphics[scale=0.4]{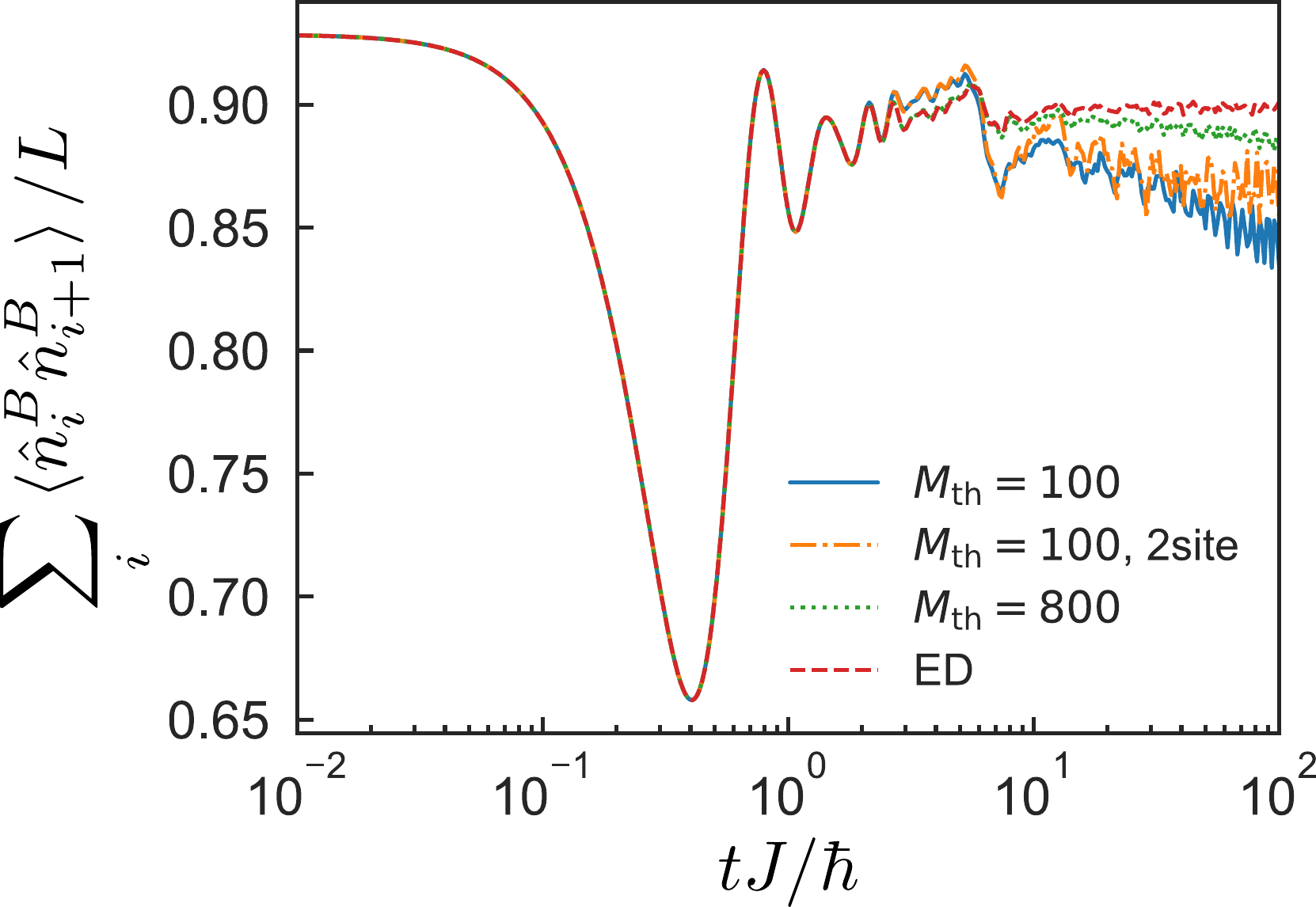}
  \caption{(Color online) Time evolution of the sum of nearest-neighbor density-density correlations for the Bose-Hubbard model after the global quench from \(U/J = \infty \) to 3.01, where \(L=N=14\) and \(V/J=0\).}\label{fig:BH_nn_inf} 
\end{figure}

\begin{figure}
  \includegraphics[scale=0.4]{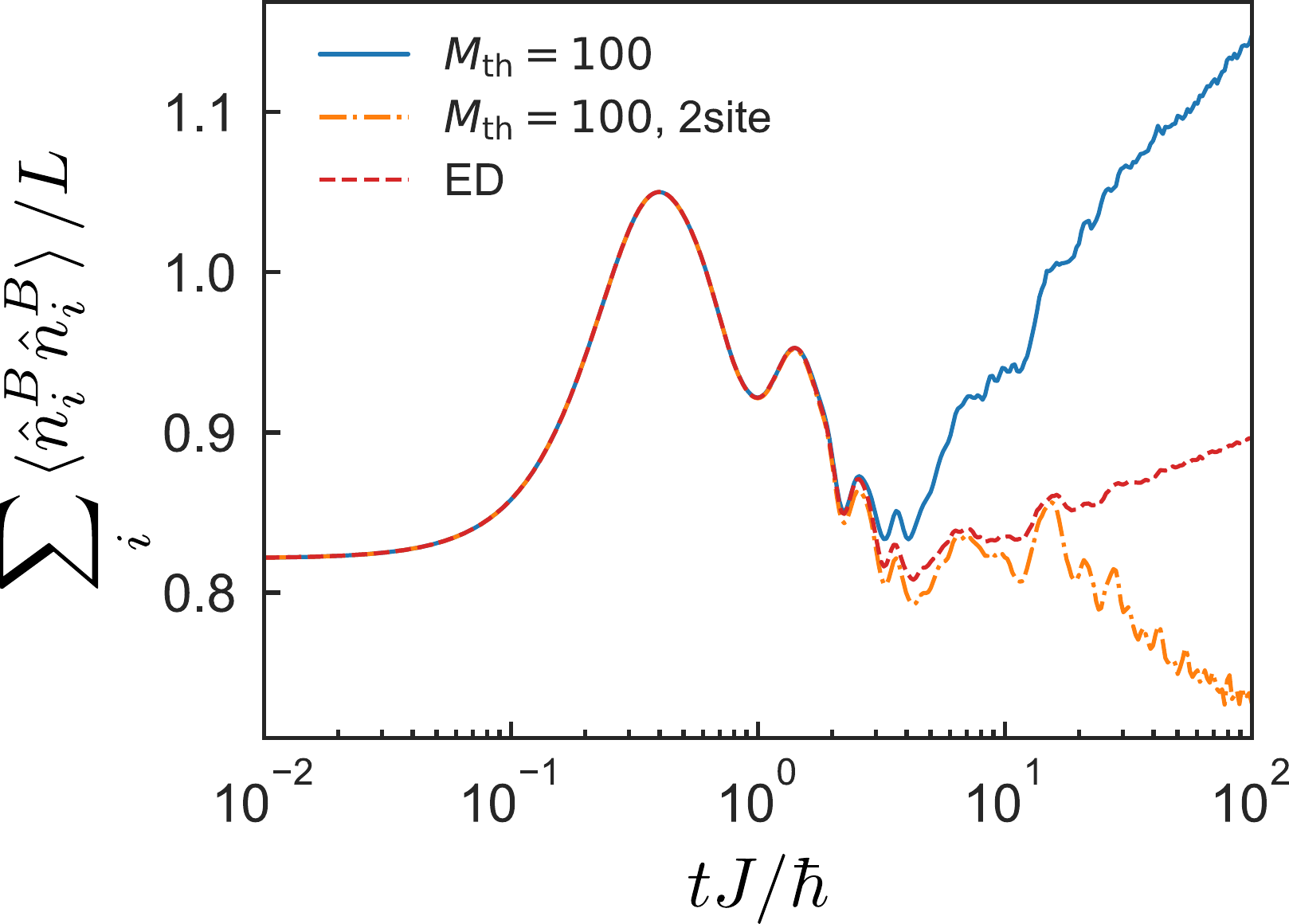}
  \caption{(Color online) Time evolution of the sum of on-site density-density correlations for the Bose-Hubbard model after the global quench from \(V/J = 0\) to 3.0, where \(U/J = 0\), \(L=20\), and \(N=10\).}\label{fig:EBH_nn_free}
\end{figure}

\begin{figure}
  \includegraphics[scale=0.4]{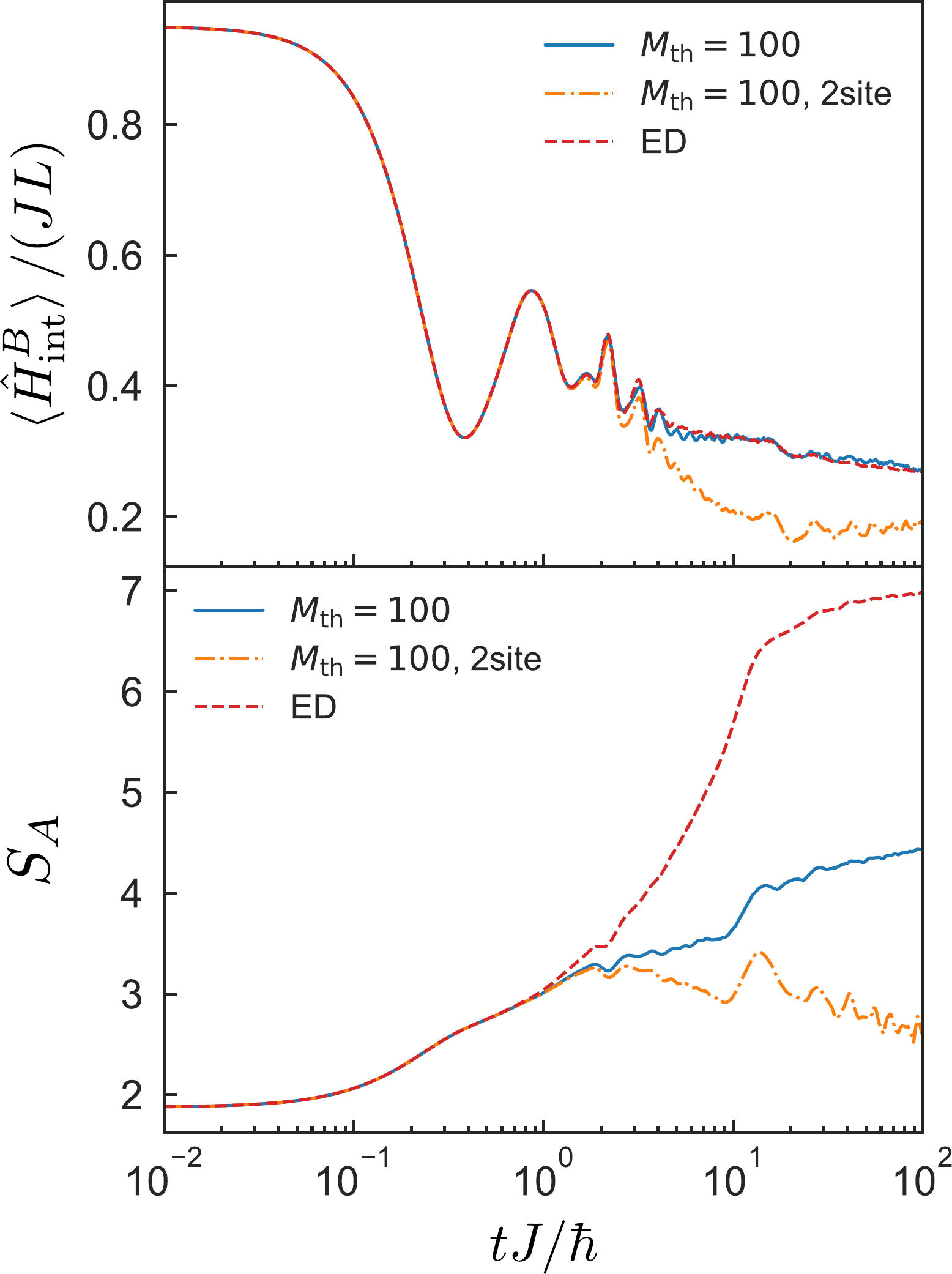}
  \caption{(Color online) Time evolution of the two quantities for the Bose-Hubbard model after the global quench from \(V/J = 0\) to 3.0, where \(U/J = 0\), \(L=20\), and \(N=10\). Upper panel: Interaction energy \(\braket{\hat{H}^B_{\rm int}}\).
  Lower panel: Entanglement entropy \(S_A\), where subsystem \(A\) is the left half of the system.}\label{fig:EBH_int_free}
\end{figure}

However, this expectation is not true.
In order to corroborate this, we depict in Fig.~\ref{fig:BH_nn_inf} the time evolution of the sum of nearest-neighbor density-density correlations \(\sum_i \hat{n}^B_i \hat{n}^B_{i+1}\)
 in the same dynamics as in Figs.~\ref{fig:BH_ene_inf} and~\ref{fig:BH_tot_inf}. 
There we see that the superiority of the hybrid scheme is absent, or rather, the two-site integration scheme gives slightly closer values to the exact ones.
A more pronounced example can be observed in the dynamics of the Bose-Hubbard model with unrealistic parameters: \(U/J = 0\) and \(V/J\) is finite.
Figure~\ref{fig:EBH_nn_free} shows the time evolution of the sum of onsite density-density correlations \(\sum_i \hat{n}^B_i \hat{n}^B_{i}\) after the global quench from \(V/J = 0\) to 3.0.
The system size \(L\) and the total number of particles \(N\) are set to 20 and 10.
Comparing the data obtained by the hybrid and two-site integration schemes, the error in the former scheme is noticeably larger than that in the latter.
It should be stressed here that when \(V/J = 0\) and \(U\) is finite, this quantity gives the interaction energy, which was well described by the hybrid scheme as shown in Figs.~\ref{fig:BH_ene_inf} and~\ref{fig:BH_ene_free}. 
In the upper panel of Fig.~\ref{fig:EBH_int_free}, we depict the interaction energy \(\braket{\hat{H}^{B}_\mathrm{int}}\) in the \(U/J=0\) system, which corresponds to the sum of nearest-neighbor density-density correlations.
We see that for the interaction energy the hybrid scheme is more accurate than the two-site scheme.
The lower panel of Fig.~\ref{fig:EBH_int_free} shows the time evolution of the entanglement entropy \(S_A\), where we see that a time-evolved state given by the hybrid scheme is more entangled than that given by the two-site integration scheme likewise \(V/J=0\) cases as shown in the lower panel of Fig.~\ref{fig:EBH_int_free}.

From these observations, we conjecture that the hybrid scheme provides more accurate results than those given by time-evolution schemes with severe truncations, such as the two-site TDVP and TEBD, for global observables included in the Hamiltonian and the entanglement entropy.
However, this superiority does not mean that a time-evolved state given by the hybrid scheme is more accurate because the hybrid scheme can be worse for other quantities.
In other words, the projection of the TDVP for MPS biases a time-evolved state towards better describing terms closely related to the total energy.

\section{Summaries\label{sec:summary}}
We studied long-time dynamics of Hubbard-type models after a sudden quantum quench in order to evaluate the performance of the time-dependent variational principle (TDVP) for a matrix product state (MPS) that circumvents increasing bond dimensions of the MPS by projecting the Hamiltonian to the manifold of MPS.\@
In the case of nonintegrable models, comparison with the numerical data obtained by the exact diagonalization indicates the superiority of the TDVP method over integration methods with the truncation of states for describing long-time behaviors of global observables included in the Hamiltonian, such as the total interaction energy.
Since the time evolution of these observables has been measured in recent experiment with ultracold atomic gases in optical lattices~\cite{takasu_experimental_2018,Asaka_JPS,Takasu_in_prepare}
, the superiority is useful for analyzing or simulating such experiments.
For an integrable model, this superiority is absent since the local update nature of TDVP breaks the conservation of the integrals of motion.
Even in nonintegrable models, we showed that the projection can cause larger error than that caused by the truncation of states for observables which are not included in the Hamiltonian.
These results mean that the projection and the energy conservation of the TDVP do not necessarily improve a time-evolved state.

\begin{acknowledgments}
We thank E\@. V\@. H\@. Doggen and E\@. M\@. Stoundenmire for useful comments.
The MPS calculations in this work are performed with ITensor library, http://itensor.org.
This work was supported by KAKENHI from Japan Society for the Promotion of Science, Grant Numbers 18K03492, 18H05228, MEXT Q-LEAP, and CREST from Japan Science and Technology Agency No.~JPMJCR1673.
\end{acknowledgments}
\bibliographystyle{apsrev4-1} 

\end{document}